\documentstyle[preprint,aps,epsf]{revtex} 
\begin{document}
\draft
\preprint{$\scriptstyle{\rm UMHEP-450} \atop{\scriptstyle
{{\rm JHU-TIPAC-98001}}}$}
\tightenlines
\title{Can Nearby Resonances Enhance $D^0 - {\bar D}^0$ Mixing?}
\author{Eugene Golowich} 
\address{Department of Physics and Astronomy,
University of Massachusetts \\
Amherst MA 01003 USA}
\author{Alexey A. Petrov}
\address{Department of Physics and Astronomy, 
The Johns Hopkins University \\
Baltimore, MD 21218 USA} 
\maketitle
\begin{abstract}
\noindent
We study the contributions of resonances to 
$D^0 - {\bar D}^0$ mixing.  Both $Q{\bar Q}$ and hybrid 
$Q{\bar Q}G$ states are considered.  
Assuming reasonable values for the resonance parameters, 
we find relatively sizeable individual contributions 
to both $\Delta m_D$ and $\Delta \Gamma_D$.
We derive a variant of the GIM cancellation mechanism
for the resonance amplitudes and show that
broken $SU(3)$ can allow for appreciable residual effects.  
Additional input from meson spectroscopy and
lattice gauge simulations will be needed to
improve the accuracy of these predictions.
\end{abstract}

\section{Basic Approach} 
In this paper, we explore the possible enhancement in the
mixing rate of neutral $D$ mesons due to nearby resonances.
It is interesting that the resonance mechanism, if operative,
can be available only to $D$ mesons.  The light kaon lies below
the resonance region and the heavy $B_{d,s}$ mesons lie above it.
The dynamical mechanism of resonant enhancement constitutes 
an explicit violation of the quark-hadron duality assumption and
could influence power counting rules built into the HQET estimate
of $\Delta m_D$~\cite{Ge92,Oh93}, which assumes a large energy gap
between $m_c$ and the scale $\Lambda_{\rm QCD}$ at which hadron
dynamics is active. In addition, if a resonance is viewed as a 
single-particle intermediate state, then its contribution will be 
favored over that of multibody intermediate states by the 
$1/N_c$ counting rules.  

For the remainder of this section, we continue the discussion of 
resonance contributions to mixing amplitudes.  
Section II concerns applications to $D^0-{\bar D}^0$ mixing, 
Section III addresses the issue of SU(3) multiplet structure and 
conclusions are presented in Section IV.

From standard perturbation theory, the $ij^{\rm th}$ element of the 
$D^0 - \bar D^0$ mass matrix can be represented as
\begin{equation}
\left [ M - i \frac{\Gamma}{2} \right ]_{ij} =
\frac{1}{2 m_D}
\langle D^0_i | {\cal H}^{\Delta C=2}_W | D^0_j \rangle
+ \frac{1}{2 m_D} \sum_{I}
\frac{\langle D^0_i | {\cal H}^{\Delta C=1}_W | I \rangle
\langle I | {\cal H}_W^{\Delta C=1 \dagger} | D^0_j \rangle}
{m_D^2 - m_I^2 + i \epsilon} \ \ . 
\label{mixmatr}
\end{equation}
The first term in the mass matrix expansion of Eq.~(\ref{mixmatr})
corresponds to the contribution of local $\Delta C = 2$ box and
dipenguin operators.  These are small in the Standard
Model~\cite{Da85,dght,ap}.  Next come the bilocal contributions
which are induced by the insertion of two $\Delta C=1$ operators.
This class of terms might be enhanced by various nonperturbative  
effects, and therefore is of considerable interest. As follows from
Eq.~(\ref{mixmatr}), one introduces a sum over all possible
$n$-particle intermediate states allowed by the corresponding
quantum numbers.  For these continuum contributions, the summation in
the second term of Eq.~(\ref{mixmatr}) takes the form of an integral 
over the energy variable.  There will be a unitarity cut in the
complex energy plane lying along the real axis and beginning at the 
two-pion threshold.  The contribution from charged pseudoscalar 
two-body intermediate states was originally considered in
Refs.~\cite{dght,lw} and estimated to be potentially large.  However, 
it remains very difficult to reliably determine the total effect
associated with $n\ge 2$ intermediate states due to the many decay
modes present, each having unknown final state interaction (FSI)
phases.  For a recent attempt in this direction, see Ref.~\cite{Bu95}.  

There are also the `single-particle' effects arising both 
from bound states and from resonance intermediate states.
A bound state contribution occurs as a pole on the real-$E$ axis.  
A resonance contribution lies in the continuum and corresponds
to a pole on an unphysical Riemann sheet.  Its contribution
will be a sharply peaked lorentzian profile of the discontinuity
across the unitarity cut, much like that of a bound state.
However, a resonance will contribute both to $\Delta m_D$ and $\Delta
\Gamma_D$.  In principle, single-particle effects are rather simpler 
to analyze.  The number of such intermediate states is constrained, 
and they can generally be estimated, at least roughly.  As mentioned 
above the $1/N_c$-counting rules, shown already to work reasonably
well for the estimates of $D$-meson decay widths, favor a set of
single-particle intermediate states, {\it i.e.} pole 
diagrams ({\it cf} Fig.~1) for self-energies of $D_{^L_S}$. 
The light-meson and $B$-flavored meson single-particle contributions to 
$D^0 - {\bar D}^0$ mixing have already been analyzed~\cite{eg97}.  
In this paper we study the resonance sector.  

Consider the special role of resonances. To begin, 
we express the collection of such contributions to $\Delta m_D$ 
(upon neglecting CP-violation) as
\begin{equation}
\Delta m_D \bigg|^{\rm res}_{\rm tot} = 
\frac{1}{2 m_D} \sum_{R} Re~
\frac{\langle D_L | {\cal H}_W | R \rangle
\langle R | {\cal H}_W^\dagger | D_L \rangle}
{m_D^2 - m_R^2 + i \Gamma_R m_D}
\ \ - \ \ (D_L \to D_S) \ \ .
\label{deltam}
\end{equation}
The pseudoscalar $0^{-+}$ (scalar $0^{++}$) intermediate states have 
$CP = -1$ ($CP = +1$) and contribute to the $D_L$ ($D_S$) part 
of the above.  If the mass of the resonance is not too far 
from the $D$-meson mass, an interesting effect occurs.  
To highlight the dependence on the resonance mass, temporarily
consider just the energy denominator in Eq.~(\ref{deltam}).
The contribution to the energy denominator of a light bound state
({\it e.g.} pions or kaons) of mass $m$ is
$P(m^2) \equiv 1/(m_D^2 - m^2) = m_D^{-2} + {\cal O}(m^2/m_D^2)$, 
which amounts to a suppression factor of order $m_D^{-2}$.  By
contrast, the energy denominator ({\it cf} Eq.~(\ref{deltam})) 
for a resonance of mass $m_R$ and width $\Gamma_R$ will yield 
\begin{equation}
\Delta m_D\bigg|^{\rm res}_{\rm R}
\propto { m_D^2 - m_R^2 \over
(m_D^2 - m_R^2)^2 + \Gamma_R^2 m_D^2}\ ,
\qquad \Delta \Gamma_D\bigg|^{\rm res}_{\rm R}
\propto - { \Gamma_R m_D \over
(m_D^2 - m_R^2)^2 + \Gamma_R^2 m_D^2} \ \ .
\label{res2}
\end{equation}
In the limit of a narrow resonance width, the 
expression for $\Delta \Gamma_D$ becomes proportional to the delta 
function $\delta(m_R - m_D)$, as expected.  The effect of a 
finite width is to allow the resonance to contribute at 
values $m_R \ne m_D$.  The contribution to $\Delta m_D$
vanishes at $m_R^2 = m_D^2$ since it undergoes a change
of sign there.  Considered as a function
of the resonance mass $m_R$, the maximum effect 
occurs for $m_R^2 = m_D^2 \pm \Gamma_R m_D$ at which
${\cal R}e ~P (m_D^2 \pm \Gamma_R m_D) = \mp 1/(2 \Gamma_R  
m_D)$.  On the other hand, the resonance contribution to
$\Delta \Gamma_D$ is maximized at the different value 
$m_R^2 = m_D^2$, but is still substantial 
at the values $m_R^2 = m_D^2 \pm \Gamma_R m_D$ which
maximize the contribution to $\Delta m_D$.  In particular, 
for both $\Delta m_D$ and $\Delta \Gamma_D$,
the $m_D^{-2}$ dependence which would appear
for a very light resonance has been replaced by
$\Gamma_R m_D$. Thus, for a resonance
sufficiently near the $D$ meson the possibility exists for
an enhancement factor of order $m_D/ \Gamma_R \simeq 5 \to 15$ for
both $\Delta m_D$ and $\Delta \Gamma_D$ relative to an
unenhanced pole contribution.  Actually it even makes sense to 
broaden the term `nearby resonance' to include $m_R > 
1$~GeV since the $m_D^{-2}$ suppression 
mentioned above will be largely overcome.  

Also present in Eq.~(\ref{deltam}) are the $D$-to-$R$ transition 
amplitudes.  To obtain a quantitative description, we shall 
adopt as our $\Delta C=1$ hamiltonian the phenomenological 
Bauer-Stech-Wirbel effective operator~\cite{bsw},
\begin{equation}
{\cal H}_W^{\rm BSW} =
{G_F a_2 \over \sqrt{2}}
\bar u_k \Gamma^\mu_{\rm L} c_k  
\Biggl [V_{cd}^* V_{us} {\bar d}_j  \Gamma_\mu^{\rm L} s_j +
V_{cs}^* V_{ud} {\bar s}_j  \Gamma_\mu^{\rm L} d_j
+ V_{cd}^* V_{ud} {\bar d}_j  \Gamma_\mu^{\rm L} d_j +
V_{cs}^* V_{us} {\bar s}_j  \Gamma_\mu^{\rm L} s_j \Biggr ] \ \ ,
\label{hbsw}
\end{equation}
where the constant $a_2 = -0.55 \pm 0.1$ is fixed from fitting
nonleptonic $D$ decays and it is understood that the operator
of Eq.~(\ref{hbsw}) is to be evaluated in vacuum saturation.  
In vacuum saturation, the contribution of resonance $R$ to mixing will be 
proportional to the squared decay constant $f_R^2$.  This has two 
important consequences:
\begin{enumerate}
\item $Q{\bar Q}$ resonances having $J^P = 0^+$ will not contribute, 
as they occur in P-waves and thus have vanishing wave function at 
the origin.  Although they could well be nonzero in a more general 
setting, their absence here suggests they would be suppressed.
\item We interpret $J^P = 0^-$ $Q{\bar Q}$ resonances with masses 
nearest the $D$ as second radial excitations.  We include also 
{\it first} radial excitations in our study due to their larger 
decay constants but omit radial excitations above the second. 
\end{enumerate}

\section{Contributions of Individual Resonances}
The fact that the most recent Particle Data Group 
compilation~\cite{Pd96} cites many nonstrange and strange 
resonances in the mass region up to $2100$~MeV strongly supports
the premise of a resonance mechanism.  Since experimental
data are still relatively sparse in the mass region of the $D$ system, 
however, additional spectroscopic knowledge of this energy range is 
needed.  

In the following, we shall consider the effects of individual 
$Q {\bar Q}G$ or $Q \bar Q$ composites.  Although individual 
contributions like these will, at least to some extent, be subject 
to GIM cancellations as other states are added in, such contributions
nonetheless serve as useful indicators of what mixing signal to be 
reasonably expected.  We employ Eq.~(\ref{hbsw}) for the $Q {\bar Q}$ 
examples, while employing a largely phenomenological method for the 
$Q {\bar Q}G$ case.

\vspace{0.2cm}

\noindent{\bf $Q \bar Q$ Resonance}

The mixing amplitudes induced by resonance $R$ are 
\begin{equation}
\Delta m_D^{(R)} = - C 
f_R^2 ~{\mu_R (1 - \mu_R) 
\over (1 - \mu_R)^2 + \gamma^2_R}\ , 
\qquad 
\Delta \Gamma_D^{\rm res} = - C 
f_R^2 ~{\mu_R \gamma_R 
\over (1 - \mu_R)^2 + \gamma^2_R} \ \ ,
\label{delmg}
\end{equation}
where $C \equiv 2 m_D (G_F a_2 f_D \xi_d /\sqrt{2})^2$, 
the dimensionless quantities 
$\mu_R \equiv m_R^2/m_D^2$ and $\gamma_R \equiv 
\Gamma_R/m_D$ are the reduced squared-mass and width 
of the resonance, and from the unitarity of CKM 
matrix and neglecting $\xi_b$ we have used $\xi_s \simeq -\xi_d$
where $\xi_i \equiv V_{ci}^* V_{ui} \ (i=d,s)$.

Since decay widths $\Gamma \simeq  0.20$~GeV are characteristic 
of resonances in the $1\to 2$~GeV mass range, one expects that 
$1 \gg \gamma$ in applications of Eq.~(\ref{delmg}).  We note that 
the two mass values $m_{\rm res}({\rm GeV}) = 1.772, 1.973$ 
maximize the resonance contribution to $D^0 - {\bar D}^0$ 
mixing for fixed decay width and decay constant values.   
As to the dependence in Eq.~(\ref{delmg}) on 
the resonance decay constant one recalls $f_\pi \simeq 0.13~{\rm
GeV}$, $f_K \simeq 0.16~{\rm GeV}$ for the noncharm ground state 
and $f_D \simeq 0.2~{\rm GeV}$, $f_{D_s} \simeq 0.28~{\rm
GeV}$~\cite{Pd96,CLEO97} for the charm ground state.  
Excitations of the constituent quarks will reduce the wave function at the
origin and thus decrease the corresponding decay constant. 
For $Q \bar Q$ radial excitations, we use hydrogen atom 
wave functions to provide a rough guide in estimating 
default values of $f_R$.  For first radial excitations, 
we estimate $f_R \simeq 0.025$~GeV whereas for 
second radial excitations, we use $f_R \simeq 0.01$~GeV in our numerical 
work.  We expect contributions from even higher 
radial excitations to be negligible.  In our numerical work 
results are scaled with the square of the associated decay constant 
to allow for any future departures in assumed values or 
quantum number assignments of individual states.  

We display in the Table various resonance contributions 
(assuming the $Q{\bar Q}$ description) to $\Delta m_D$ and to $\Delta 
\Gamma_D$.  The mass and decay width values are taken from the PDG
listing.  We do not intend our listing to be complete, but 
instead to indicate the magnitudes associated with 
contributions of this type.  
Although smaller than the current experimental limit~\cite{Pd96,E791} 
$|\Delta m_D|^{\rm expt} < 1.3 \cdot 10^{-14}~{\rm GeV}$, 
the values are larger than 
the contribution $\Delta m_D|^{\rm gnd~state} 
\simeq 3 \cdot 10^{-17}~{\rm GeV}$ 
from the set of pseudoscalar ground
state mesons ($\pi$, $K$, $\eta$, $\eta'$).  They also tend to 
dominate the short distance `box' contributions 
$|\Delta m_D|^{\rm box} \simeq 1.9 \times 10^{-17}~{\rm GeV}$ and 
$|\Delta \Gamma_D|^{\rm box} \simeq 0.75 \times 
10^{-17}~{\rm GeV}$, where we have taken $m_c = 1.3$~GeV, 
$m_s = 0.2$~GeV, and refer effects of QCD radiative 
corrections to Ref.~\cite{Oh93}.

\vspace{0.2cm}

\noindent{\bf $Q {\bar Q}G$ Resonance}

In vacuum saturation, contributions to $\Delta m_D$ from 
$Q{\bar Q}$ intermediate states arise mainly from annihilation amplitudes.
It is well known that such amplitudes are subject to helicity  
suppression.  This is the same effect which influences the observed 
patterns of leptonic pion and kaon decay modes.  It was noted long 
ago that the effects of helicity supression can be (partially)
lifted by soft-gluon emission.  While this mechanism
can be readily applied to inclusive heavy meson decay, it
is difficult to see how to implement it to the mixing
matrix elements of Eq.~(\ref{deltam}) if the intermediate states
are of the $Q{\bar Q}$ variety.  However, this difficulty is
avoided if the intermediate state meson is a $Q \bar Q G$ hybrid
state which involves a constituent gluon. The argument can be
further extended to include penguin operators, thus introducing a
long-distance counterpart of the dipenguin operator
contribution~\cite{ap,aapbnl}. It is plausible to assume that the gluon 
produced by a penguin (or any other operator) can form a
quasibound state along with the $u{\bar u}$ quark-antiquark pair,
thus producing a hybrid resonance ({\it cf} Fig.~2). 
These one-particle intermediate state
contributions to $\Delta m_D$ might be of importance
because of the proximity of the anticipated~\cite{lip,cl}
hybrid-meson mass with that of the $D$-meson.

In this regard, a particularly interesting candidate is
the $\pi_H (1800)$ which has $J^{PC} = 0^{-+}$.  On the
basis of reports from
several experimental groups~\cite{exp} along with various
quark model analyses~\cite{theor}, it is tempting to assign
this particle as a hybrid.  One can then estimate the contribution
of $\pi_{\rm H}(1800)$ to Eq.~(\ref{deltam}) provided
that the mixing amplitude 
$g \equiv \langle D_L | H_w | \pi_{\rm H}(1800) \rangle$ is known.
This amplitude can be inferred from quark models or even better,
determined phenomenologically by using available data on
$D$ decay rates. The idea is to search for common decay channels of
$D$ and $\pi_{\rm H}(1800)$ where the $\pi_{\rm H}(1800)$
contribution is manifest
and then estimate the mixing amplitude from this.  The situation
is as depicted in Fig.~3, in which $D-\pi_{\rm H}(1800)$
mixing is followed by $\pi_{\rm H}(1800)$ decay.

It was noted in
theoretical calculations~\cite{theor} and hinted at experimentally that
the decay rates $\pi_{\rm H}(1800) \to \pi f_0(980),~
\pi f_0(1300)$ are large for a hybrid
$\pi_{\rm H}(1800)$.
Thus one can put an upper bound on the
mixing amplitude $g$ by introducing a model for the resonant
decay of $D$-meson via $\pi_{\rm H}(1800)$~\cite{E687},
\begin{equation}
{\cal M}_{D \to \pi f_0(980)} = 
{g \over m_D^2 - m_{\pi_{\rm H}}^2 + i \Gamma_{\pi_{\rm H}} m_D} 
{\cal M}_{\pi_{\rm H} \to \pi f_0(980)} \ \ .
\label{aapmodel}
\end{equation}
The partial decay width for $\pi_H \to \pi f_0(980)$ can
be written as
\begin{equation}
\Gamma_{\pi_H \to \pi f_0(980)} = \frac{1}{16 \pi  
m_{\pi_H}} \left |
{\cal M}_{\pi_{\rm H} \to \pi f_0(980)} \right |^2 \lambda_\pi \ \ , 
\label{ratepih}
\end{equation}
where $\lambda_I$ for $I \to f_1 f_2$ is defined as
\begin{equation}
\lambda_I^2 \equiv 
\bigg[ 1 -
{(m_{f_1} + m_{f_2})^2 \over m_I^2} \bigg] \bigg[ 1 -
{(m_{f_1} - m_{f_2})^2 \over m_I^2} \bigg] \ .
\label{lamdef}
\end{equation}
A similar formula exists for the $D \to \pi f_0(980)$ transition.
In the simplest case, the total decay width of $\pi_H$ can be
saturated by the single partial decay width
$\Gamma_{\pi_H \to \pi f_0(980)}$.
This is a reasonable approximation as this decay mode becomes  
dominant for
the hybrid $\pi_H$. Thus, using Eq.~(\ref{ratepih}) along with
an expression for $\Gamma_{D \to \pi f_0(980)}$, the mixing
amplitude $g$ can be estimated from
\begin{equation}
|g|^2 =
{m_D \Gamma_D \lambda_\pi
\over m_{\pi_H} \Gamma_{\pi_H} \lambda_D } \cdot
\left[ (m_D^2 - m_{\pi_H}^2)^2 +
\Gamma_{\pi_H}^2 m_D^2 \right] \ \ . 
\label{hyest}
\end{equation}
Computing the mixing amplitude $g$ using experimental data on the
decay rate $\Gamma_{D \to \pi f_0(980)}$ and inserting it into
Eq.~(\ref{deltam}), we estimate
$|\Delta m_D|^{\pi_{\rm H}(1800)}
\leq 0.3 \times 10^{-16}~{\rm GeV}$, 
comparable to the short distance result.

\section{Effect of Multiplet Structure}
Resonances contributing as intermediate states to $D^0 - {\bar D}^0$ 
mixing are expected to occur as SU(3) flavor multiplets. The
contribution of an entire multiplet will vanish in the limit of
degenerate light-quark masses due to GIM cancellation.
It is not clear how powerful the GIM suppression will
be, as SU(3) is known to be badly broken in at least
some $D$ decays.  For example, there is the experimentally
measured ratio~\cite{Pd96,E791a}
$\Gamma_{D^0 \to K^+K^-}/\Gamma_{D^0 \to \pi^+ \pi^-} \simeq 3$,
which is unity in the $SU(3)$ limit.  Theoretically, it has been
suggested that such large breaking is an accumulation of a number
relatively minor effects whose ultimate impact is
substantial.~\cite{cc94}  At any rate, large SU(3) breaking could 
produce a loophole for evading GIM suppression.

Consider an octet of excited mesons $\pi_{\rm H}$, $K_{\rm H}$,
${\bar K}_{\rm H}$, $\eta_{\rm H}$. We use the
subscript `H' to represent {\it heavy} mesons,
and denote the individual members of a resonance octet with the
above flavor labels.  We anticipate the presence of a ninth heavy meson
$\eta'_{\rm H}$ to allow for mixing occurring with
$\eta_{\rm H}$.  In principle, the mixing angle $\theta_{\rm H}$ 
ccn be inferred from either from mass determinations 
or from two-photon branching ratios~\cite{Pd96,dgh}.

We write for the contribution of a mixed octet of resonances to
$\Delta m_D$ and $\Gamma_D$,  
\begin{eqnarray}
& & \Delta m_D|^{\rm res}_{\rm octet} = 
\Delta m_D^{(K_H)} - {1\over 4} \Delta m_D^{(\pi_H)} 
- {3 \cos^2\theta_{\rm H} \over 4} \Delta m_D^{(\eta_H)} 
- {1 \sin^2\theta_{\rm H} \over 4} \Delta m_D^{(\eta'_H)} 
\label{m-conven} \\
& & \Delta \Gamma_D|^{\rm res}_{\rm octet} = 
\Delta \Gamma_D^{(K_H)} - {1\over 4} \Delta \Gamma_D^{(\pi_H)} 
- {3 \cos^2\theta_{\rm H} \over 4} \Delta \Gamma_D^{(\eta_H)} 
- {1 \sin^2\theta_{\rm H} \over 4} \Delta \Gamma_D^{(\eta'_H)} \ \ ,
\label{g-conven} 
\end{eqnarray}
where $\Delta {m_D}^{(i)}$ and $\Delta {\Gamma_D}^{(i)}$ are as 
in Eq.~(\ref{delmg}).
The effect of SU(3) breaking can further be studied, say for $\Delta m_D$, 
by expressing the octet decay constant, mass and decay-width factors 
in Eq.~(\ref{m-conven}) as
\begin{equation}
f_i = f_0 + \delta f_i \ , \qquad
\mu_i = \mu_0 + \delta\mu_i \ , \qquad 
\gamma_i = \gamma_0 + \delta\gamma_i \qquad (i = 1,\dots 8) \ \ ,
\label{brk}
\end{equation}
where $f_0, \mu_0, \gamma_0$ and $\delta f_i$, $\delta
\mu_i$, $\delta \gamma_i$ represent respectively the SU(3)-invariant 
and SU(3)-breaking components.  This allows for the possibility
that the result will be generally influenced by SU(3)-breaking in
the decay constant, mass and decay-width sectors.  An expression valid
to first order in symmetry breaking is
\begin{eqnarray}
& & \Delta m_D\bigg|_{\rm SU(3)~brk} = - 
{C\over 4} \bigg[ F_f (f_0,\mu_0,\gamma_0)\left(
4 \delta f_{K_{\rm H}} - \delta f_{\pi_{\rm H}}
- 3 \delta f_{\eta_{\rm H}} \right)  
\nonumber \\
& & + F_\mu (f_0,\mu_0,\gamma_0)\left(
4 \delta \mu_{K_{\rm H}} - \delta \mu_{\pi_{\rm H}}
- 3 \delta \mu_{\eta_{\rm H}} \right)
+ F_\gamma (f_0,\mu_0,\gamma_0)\left(
4 \delta \gamma_{K_{\rm H}} - \delta \gamma_{\pi_{\rm H}}
- 3 \delta \gamma_{\eta_{\rm H}} \right) \bigg] \ \ ,
\label{brk1}
\end{eqnarray}
where for simplicity $\eta_{\rm H}$-$\eta_{\rm H}'$ mixing is ignored 
and we define
\begin{equation}
F_f \equiv f_0 {2\mu_0 (1 - \mu_0) \over
(1 - \mu_0)^2 + \gamma_0^2} \ , \quad 
F_\mu \equiv f^2_0 { (1 - \mu_0)^2 + (1 - 2 \mu_0)\gamma_0^2
\over \left[(1 - \mu_0)^2 + \gamma_0^2\right]^2} \ , \quad
F_\gamma \equiv - f^2_0 {2\gamma_0 \mu_0 (1 - \mu_0) \over
\left[(1 - \mu_0)^2 + \gamma_0^2\right]^2} \ .
\label{brk2}
\end{equation}
Due to the relative lack of data, it is not possible at
this time to provide a unique analysis of the above relations.
Either a small or large effect could emerge, for example:
\begin{enumerate}
\item In the absence of $\eta_{\rm H}$-$\eta_{\rm H}'$ mixing,
the combination $4 \delta \mu_{K_{\rm H}}$ - $\delta \mu_{\pi_{\rm H}}$ 
- $3 \delta \mu_{\eta_{\rm H}}$ vanishes by virtue of the
Gell~Mann-Okubo formula and the remaining dependence in
Eq.~(\ref{brk1}) vanishes with the choice $\mu_0 = 1$
(implying $F_f = F_\gamma = 0$).
\item
The choice $\mu_0 = 1 - \gamma_0$ (with $1 \gg \gamma_0$) yields
$$
{F_f \simeq {f_0 \over \gamma_0}\ ,
\quad F_\mu \simeq {f_0^2 \over 2\gamma_0}\ ,
\quad F_\gamma \simeq - {f_0^2 \over 2\gamma^2_0}} \ .
$$
Here the net effect appears in terms of fractional changes
in decay constants and decay rates,
$$ 
|\Delta m_D|^{\rm SU(3)~brk} \simeq
0.2 \times 10^{-14}{f_0^2 \over \gamma_0}~{\rm GeV} 
\bigg| {4 \delta f_{K_{\rm H}} - 
\delta f_{\pi_{\rm H}} - 3 \delta f_{\eta_{\rm H}} \over f_0}
- {4 \delta \gamma_{K_{\rm H}} - \delta \gamma_{\pi_{\rm H}}
- 3 \delta \gamma_{\eta_{\rm H}} \over 2 \gamma_0} \bigg| \ \ .
$$
\end{enumerate}

In the first of the above items, the vanishing of
$\Delta m_D|^{\rm SU(3)-brk}$
to first order in SU(3) symmetry breaking occurs for
a special parameter choice and is clearly more an
exception than a rule.  The second item has the SU(3) degenerate
mass set at a lower value (still with no 
$\eta_{\rm H}$-$\eta_{\rm H}'$ mixing) and a nonzero effect
will generally occur, although to be more quantitative would require 
additional experimental input.  The presence of large symmetry 
breaking might necessitate a treatment beyond the first-order
relations given above.
\section{Concluding Remarks}
The motivation most often cited in searches
for $D^0 - {\bar D}^0$ mixing lies with the possibility of
observing a signal from new physics which dominates that
from the Standard Model.  The best experimental limit,
recently obtained by E791~\cite{E791},
is well beneath existing estimates of the Standard Model
value.  There are plans to improve on the E791 determination,
both at B-factories \cite{gb} and at hadron colliders \cite{ss97}.  
In addition, preliminary plans at Jefferson Lab to build
a new experimental hall and simultaneously to 
raise the beam energy suggest the possibility 
for $D^0 - {\bar D}^0$ mixing studies at that facility as well.  
For all such efforts, it will be crucial to understand the
magnitude of the mixing amplitude from the Standard Model.

In this paper, we have studied the set of potentially significant 
contributions to $D^0 - {\bar D}^0$ mixing from 
pseudoscalar resonances.  We have shown how an enhancement
of order $m_D / \Gamma_R$ can arise from a resonance whose mass 
lies within several decay widths of $m_D$ and have 
also have pointed out the importance of lighter resonances 
due to decay constant dependence in the mixing amplitude.  
In order to obtain a more detailed 
understanding of the resonance scenario, we have considered
contributions from both traditional ${\bar Q}Q$ resonances 
as well as an exotic ${\bar Q}QG$ hybrid.  Effects of order 
$10^{-16}$~GeV are possible for both $\Delta m_D$ and 
$\Delta \Gamma_D$.  In addition, it would appear possible 
or even likely in the resonance mechanism that 
$|\Delta \Gamma_D / \Delta m_D| \simeq 1$ or even larger.
This calls into question the usual assumption, 
that $\Delta m_D \gg \Delta \Gamma_D$, 
made in searches for CP violation in D-decay using
time-dependent measurements~\cite{Bl95}.  We shall consider 
generalizations of our approach and implications of our findings 
{\it vis-a-vis} CP-violating signals in a separate publication.  

Of course, efforts such as this are severly hampered by a lack of
knowledge regarding the properties of mesons lying in the
$1.6 \to 2.1$~GeV mass range.  Two
kinds of additional input would be of significant value to this
subject.  From experimentalists could come a more
complete listing of resonance mass and decay-width parameters.  
Information about various decay modes could
allow a distinction between the ${\bar Q}Q$ and ${\bar Q}QG$
descriptions.  The lattice-gauge community could supply 
numerical estimates of both decay constants of excited mesons and 
also, as a test of vacuum saturation,  
matrix elements like $\langle a_0 | H_W | D^0 \rangle$.   
As such valuable information becomes 
available, we can anticipate real progress in this area.
\section{Acknowledgments}
We would like to thank J. Donoghue, A. Falk, N. Isgur, R. Lewis, 
P. Page, and A. Zaitsev for useful conversations.
The research described in this paper was supported in part by
grants NSF PHY-9218396, NSF PHY-9404057, NSF PHY-9457916
from the National Science Foundation and by grant
DE-FG02-94ER40869 from the Department of Energy.




\begin{figure}[htb]
\epsfxsize 3.0in
\centerline{
\epsfbox{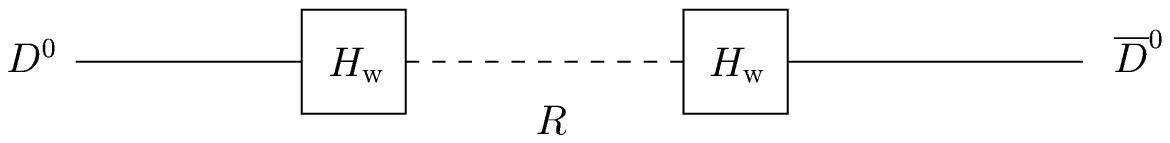}}
\caption{Contribution of resonance $R$ to the
$D^0$-to-${\bar D}^0$ matrix element.}
\label{fig:resR}
\end{figure}

\begin{figure}[htb]
\centerline{
\epsfxsize 2.0in
\epsfbox{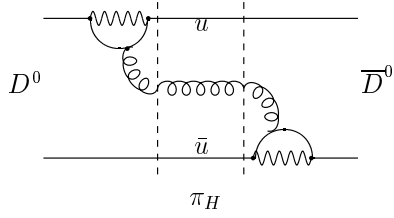}}
\caption{$Q {\bar Q}G$ intermediate state.}
\label{fig:hybrid}
\end{figure}

\begin{figure}[htb]
\centerline{
\epsfxsize 3.0in
\epsfbox{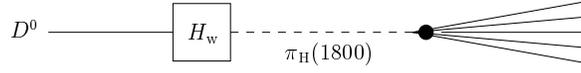}}
\caption{Hybrid contribution to $D^0$ decay.}
\label{fig:model}
\end{figure}

\begin{table}
\begin{tabular}{lll}
\multicolumn{3}{c}{Table: {Magnitudes of Pseudoscalar Resonance 
Contributions.}} \\
\hline\hline
Resonance & $|\Delta m_D| \times 10^{-16}$~(GeV) & 
$|\Delta \Gamma_D| \times 10^{-16}$~(GeV)  \\ \hline\hline
$K (1460)$ & $\sim 1.24~(f_{K(1460)}/0.025)^2$ & 
$\sim 0.88 ~(f_{K(1460)}/0.025)^2$ \\
$\eta (1760)$ & $(0.77 \pm 0.27)~(f_{\eta(1760)}/0.01)^2$ &
$(0.43 \pm 0.53)~(f_{\eta(1760)}/0.01)^2$ \\
$\pi (1800)$ &  $(0.13 \pm 0.06)~(f_{\pi(1800)}/0.01)^2$ & 
$(0.41 \pm 0.11)~(f_{\pi(1800)}/0.01)^2$ \\
$K (1830)$ & $\sim 0.29 ~(f_{K(1830)}/0.01)^2$ & 
$\sim 1.86 ~(f_{K(1830)}/0.01)^2$ \\
\hline
\end{tabular}
\end{table}

\end{document}